%% file: main.tex
\documentclass[sigconf]{acmart}
\AtBeginDocument{%
  }

\copyrightyear{2026}
\acmYear{2026}
\setcopyright{cc}
\setcctype{by-nc-nd}
\acmConference[SIGMOD Companion '26]{Companion of the International Conference on Management of Data}{May 31-June 05, 2026}{Bengaluru, India}
\acmBooktitle{Companion of the International Conference on Management of Data (SIGMOD Companion '26), May 31-June 05, 2026, Bengaluru, India}
\acmDOI{10.1145/3788853.3803098}
\acmISBN{979-8-4007-2450-3/2026/05}

\usepackage{cleveref}
\usepackage{listings}
\usepackage{stfloats}
\usepackage{subcaption}
\usepackage{balance}

\input{macros}

\begin{document}

\title{Enzyme: Incremental View Maintenance for Data Engineering}

\author{Ritwik Yadav}
\authornote{Coauthored this paper.}
\orcid{0009-0000-9428-5884}

\author{Supun Abeysinghe}
\authornotemark[1]
\orcid{0000-0001-6054-2432}

\author{Min Yang}
\authornotemark[1]
\orcid{0009-0006-2911-0985}

\author{Jeffrey Helt}
\authornotemark[1]
\orcid{0000-0003-1192-7111}

\author{Manuel Ung}
\authornotemark[1]
\orcid{0009-0000-7565-9203}

\author{Yuhong Chen}
\authornotemark[1]
\orcid{0009-0008-5149-7228}

\affiliation{%
  \institution{Databricks, Inc.}
  \city{San Francisco}
  \state{CA}
\country{USA}}

\author{Melody Hu}
\orcid{0009-0009-1018-3991}

\author{William Wei}
\orcid{0009-0004-4758-4709}

\author{Yiming Yang}
\orcid{0009-0004-3362-0596}

\author{Tom van Bussel}
\orcid{0009-0002-4490-3466}

\author{Sourav Chatterji}
\orcid{0009-0004-2525-4506}

\author{Indrajit Roy}
\orcid{0009-0000-2275-6640}

\affiliation{%
  \institution{Databricks, Inc.}
  \city{San Francisco}
  \state{CA}
\country{USA}}

\author{Paul Lappas}
\authornotemark[2]
\orcid{0009-0009-0029-5172}

\author{Yannis Papakonstantinou}
\authornote{Work done while at Databricks.}
\orcid{0009-0007-6360-9496}

\affiliation{%
  \institution{Google, Inc.}
  \city{Mountain View}
  \state{CA}
\country{USA}}

\author{Tahir Fayyaz}
\orcid{0009-0002-9247-781X}

\author{Bilal Aslam}
\orcid{0009-0005-4337-0575}

\author{Ross Bunker}
\orcid{0009-0004-7136-9216}

\author{Michael Armbrust}
\orcid{0009-0007-5426-3681}

\author{Shrikanth Shankar}
\authornotemark[1]
\orcid{0009-0000-2398-3431}

\affiliation{%
  \institution{Databricks, Inc.}
  \city{San Francisco}
  \state{CA}
\country{USA}}
\email{enzyme-paper@databricks.com}

\renewcommand{\shortauthors}{Ritwik Yadav et al.}

\begin{abstract}
\input{sections/abstract}
\end{abstract}

\ccsdesc[500]{Information systems~Online analytical processing engines}

\keywords{Incremental view maintenance; materialized views; ETL}

\maketitle

\input{sections/introduction}
\input{sections/background}
\input{sections/ivm_definition}
\input{sections/architecture}
\input{sections/lessons}
\input{sections/experimental_study}
\input{sections/related_works}
\input{sections/conclusion}

\bibliographystyle{ACM-Reference-Format}
\balance
\bibliography{references}

\end{document}

%% file: macros.tex
\newcommand{\vendor}{CV-IVM}

\lstdefinelanguage{DBSQL}{
  language=SQL,
  morekeywords={
    SCHEDULE,
    EVERY,
    MATERIALIZED,
    VIEW
  }
}

\DeclareMathOperator{\COUNT}{COUNT}
\DeclareMathOperator{\SUM}{SUM}
\DeclareMathOperator{\AVG}{AVG}
\newcommand{\currenttimestamp}[0]{\texttt{current\_\allowbreak{}timestamp()}}
\newcommand{\currentdate}[0]{\texttt{current\_\allowbreak{}date()}}

%% file: sections/abstract.tex
Materialized views are a core construct in database systems, used to accelerate analytical queries and optimize batch pipelines for extract–transform–load (ETL) workflows. Maintaining view consistency as underlying data evolves is a fundamental challenge, especially in high-throughput and real-time settings. Incremental view maintenance (IVM) has been studied for decades and continues to attract significant investment from major database vendors. However, most industrial systems either offer limited SQL-operator coverage or require users to hand-tune refresh strategies.

This paper presents Enzyme, an IVM engine built at Databricks to power its declarative data pipelines platform.
It provides a built-in, end-to-end approach to incremental pipelines, utilizing materialized views as first-class building blocks. By automating refresh planning, Enzyme reduces total cost of ownership and allows users to focus on business logic rather than MV mechanics. Validation across thousands of large-scale production pipelines spanning diverse application domains has demonstrated substantial computational efficiency gains, yielding a cumulative daily compute reduction of billions of CPU seconds. Built atop Apache Spark primitives, Enzyme adds a cost-based optimization layer that selects refresh strategies for collections of materialized views organized into pipelines.

Enzyme’s modular architecture is designed to generalize across data sources and query engines. We present key design decisions for incremental refresh planning and execution, including optimizations that exploit batching opportunities across materialized view sources. Experimental results on standard benchmarks demonstrate significant performance improvements at scale.

%% file: sections/introduction.tex
\section{Introduction}

Materialized views (MVs) are a fundamental building block offered by most database management systems \cite{bello1998materialized, goldstein2001optimizing, zilio2004recommending, armenatzoglou2022amazon, mcsherry2022materialize}. They are used primarily for reducing query execution costs by amortizing expensive, shared computation across multiple queries as well as improving the latency of individual queries. MVs have evolved from their traditional role as query acceleration constructs to become fundamental building blocks in modern data flow architectures. Along with data streams and primitives designed to process changes to datasets, MVs enable efficient extraction, transformation, and loading (ETL) of data at scale.

\begin{figure}[tb]            
  \centering
  \includegraphics[width=.98\linewidth]{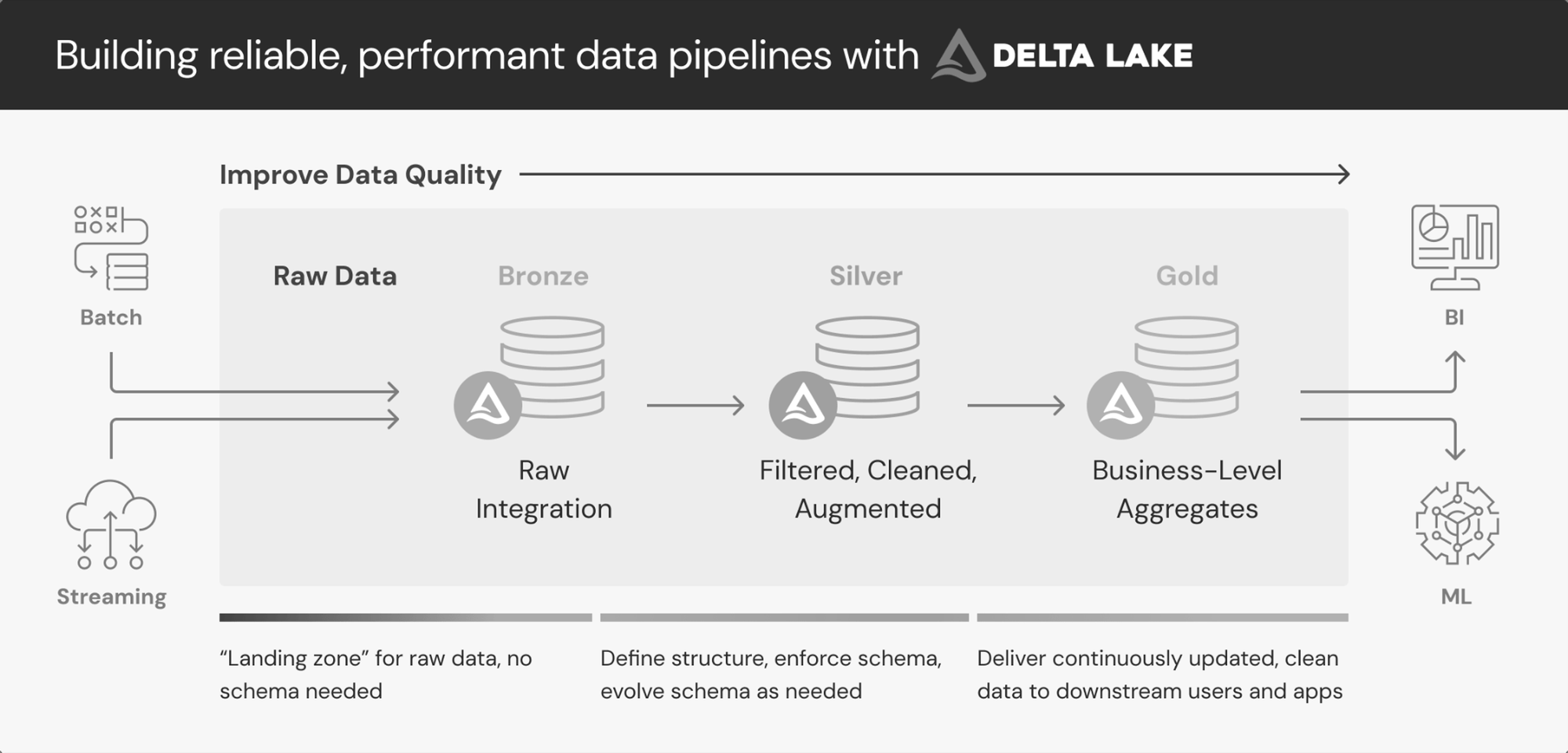} 
  \caption{Medallion architecture organizing data into progressively refined layers for analytics.}
  \Description{Figure illustrating the construction of data pipelines, with artifacts organized into progressively refined layers following the medallion architecture.}
  \label{fig:medallion}
\end{figure}

Modern data architectures, such as the medallion architecture (Figure~\ref{fig:medallion}) commonly used in data lakehouses, leverage this capability to organize and transform data across multiple layers. The quality of the data improves as it flows through subsequent layers in this architecture. 
The bronze layer lands the data as-is along with additional metadata. The silver layer matches, merges, conforms and cleans the data in the bronze layer to enable self-service analytics for ad-hoc reporting, advanced analytics, and machine learning. Finally, the gold layer performs transformations for reporting, utilizing more denormalized and read-optimized data models with fewer joins. MVs primarily feature in the silver and gold layers to apply these transformations.

Efficiently maintaining view consistency as source data evolves has been extensively studied in database theory over the last few decades \cite{gupta1993maintaining}. However, with the advent of the AI era, the scale of data at organizations has grown exponentially, making incremental refresh of MVs a necessity rather than treating it as an optional optimization. This shift has exposed several critical gaps in existing IVM systems. Modern streaming engines offer IVM as a stream processing construct, updating MVs with every change to their sources. While this approach minimizes data latency, batching changes can amortize infrastructure costs, reduce scheduling overhead, and improve resource utilization across large-scale deployments. Additionally, most prior IVM research relies on efficient secondary indexes for incremental query plans to significantly outperform full recomputation. Modern lakehouses, however, lack traditional secondary indexes and instead rely on metadata-based indexing at the file and column level. Research to reduce reliance on secondary indexes \cite{katsis2015utilizing} is relatively recent. Furthermore, comprehensive coverage for arbitrary compositions of common SQL operators---such as aggregations and window functions---remains rare in industrial-grade IVM engines.

Building a production-ready IVM engine extends far beyond constructing efficient incremental query plans. It requires complex supporting infrastructure, including:
\begin{itemize}
    \item A query fingerprinter capable of detecting changes to MV definitions across multiple languages (SQL, Python) and in the presence of user-defined functions. The diversity in languages makes it a particularly difficult problem and inaccuracies in this component can lead to incorrect results (false negative) or full recomputes on petabyte scale datasets (false positive).
    \item A cost model that estimates computational complexity of logically equivalent plans, incorporates feedback from historical refreshes, and optimizes across a DAG of MVs while providing explainable output.
    \item Specialized query mutators to support incrementalization of operators such as \currenttimestamp{} and AI functions. These operators are not addressed in traditional IVM literature but are extremely popular among data engineers and practitioners building ETL pipelines.
    \item Comprehensive correctness testing frameworks capable of non-deterministically generating test cases spanning the entire domain of valid MV definitions. The level of test coverage obtained using these frameworks cannot be matched by traditional unit or integration tests.
\end{itemize}

This paper introduces Enzyme, an out-of-the-box IVM engine developed at Databricks that addresses these challenges through automated, cost-based optimization of MV
refreshes. Enzyme provides comprehensive operator coverage, supports both scheduled batch and continuous refresh patterns, and automatically selects between incremental and full refresh strategies based on workload characteristics and historical performance. We begin by enumerating the key primitives and features that Enzyme relies upon (Section 2), followed by a formalization of the incremental view maintenance problem (Section 3). Section 4 presents Enzyme's design and architecture, while Section 5 discusses lessons learned from its implementation. We validate Enzyme on both production workloads and standard ETL benchmarks, with quantitative performance results presented in Section 6. Section 7 compares Enzyme with related systems before we conclude in Section 8.

%% file: sections/background.tex
\section{Background}

In this section, we discuss how MVs fit into the Databricks platform. We then introduce the necessary concepts and features for understanding the Enzyme incremental view maintenance system.

\subsection{Materialized Views at Databricks}
\label{sec:materialized-views-databricks}

Databricks's materialized views (MVs) physically store query results in Unity Catalog–managed tables~\cite{chandra2025unity}, in contrast to standard views that are evaluated on demand. Users can create MVs either within Spark Declarative Pipelines or through Databricks SQL (DBSQL) warehouses.
When defined inside a pipeline, an MV is refreshed whenever the pipeline itself is refreshed.
Standalone MVs, on the other hand, can be refreshed on a user-specified schedule (e.g., every hour) or triggered automatically based on changes to their input tables.
Each refresh logically updates the MV to reflect the result of re-running its defining query over the latest versions of its inputs when the refresh starts. Between refreshes, the MV can lag behind its source tables, eventually becoming consistent as subsequent refreshes incorporate newer changes. Delta Lake's ACID transactions ensure readers always see the last fully committed MV state; refreshes are either entirely visible or not at all.

\Cref{fig:running_example_mv} shows an MV that computes
the average order for three selected regions
and refreshes hourly.
We use this MV as a running example throughout the paper.

\begin{figure}[t]
\centering
\begin{lstlisting}[language=DBSQL, frame=none]
CREATE MATERIALIZED VIEW region_avg_sales
SCHEDULE EVERY 1 HOUR
AS
SELECT
  c.region,
  AVG(o.amount) AS avg_order_amount
FROM
  Customers c
  JOIN Orders o ON c.customer_id = o.customer_id
GROUP BY
  c.region
HAVING
  c.region IN ('us-east', 'us-west', 'asia');
\end{lstlisting}
\caption{
A standalone MV that computes the average order amount for three selected regions.
The MV is configured to refresh every hour.
We use this as a running example throughout the paper.
}
\label{fig:running_example_mv}
\end{figure}

\paragraph{Simplified Incrementalization and Orchestration}
MVs remove the complexity of incrementalization and
orchestration from the user.
For instance, for the MV in \Cref{fig:running_example_mv}, the underlying engine
automatically tracks changes to the input tables (\texttt{Customers} and \texttt{Orders}) and \emph{incrementally} updates the MV accordingly.
When MVs are used within a pipeline, the engine also identifies
dependencies among pipeline entities and orchestrates their refresh order
according to the underlying dependency DAG, exploiting parallelism when possible. This ensures that, for dependencies captured within the same pipeline, all downstream MVs are refreshed against a consistent, committed snapshot of their upstream dependencies.

Databricks additionally provides streaming tables with
easy-to-use declarative APIs such as \texttt{AUTO CDC}, which implement
change data capture (both SCD Type 1 and Type 2) and handle
out-of-order records~\cite{autocdc}.
These streaming tables, when paired with MVs, allow users to build large,
complex pipelines in a fully declarative manner.

Enzyme is the incremental execution engine that powers efficient materialized view refreshes in Databricks.
On each refresh, Enzyme evaluates multiple maintenance strategies (several incrementalization techniques plus the option to perform a full recompute and overwrite the MV's contents) and selects the most efficient plan.
Whenever possible, Enzyme processes only new data and changes in the source tables, eliminating the need for users to implement manual incremental logic and significantly reducing redundant computation.

\paragraph{Internal Representation of MVs}
Each MV consists of two components: a \emph{backing table} and a \emph{top-level view}. The backing table stores the user's data plus internal metadata columns. The top-level view projects away the metadata, exposing only user-requested columns.
This separation allows Enzyme to augment internal storage while maintaining a consistent external interface. For instance, as discussed later in \Cref{subsec:ivm_special_cases}, Enzyme decomposes \texttt{AVG} into \texttt{SUM} and \texttt{COUNT} \cite{gupta1993maintaining}, storing both components in the backing table but exposing only the original \texttt{AVG} column through the top-level view. \Cref{sec:decomposition} describes this and other such \emph{technique enablers} in detail, including those for supporting complex aggregations and window functions.

\subsection{Spark Primitives}

Enzyme relies on two Spark primitives to apply computed changes to materialized views:
\texttt{MERGE INTO}~\cite{mergeinto} applies updates, insertions, and deletions from a source table to a target table. It efficiently executes conditional actions by matching source and target rows on a merge condition.

\texttt{REPLACE WHERE}~\cite{replacew}, on the other hand, is simpler. Given a predicate and a set of new rows, it atomically replaces rows matching the predicate in the target table with the provided data.

\subsection{Table Format Features}

Delta Lake provides several features that Enzyme leverages for incremental maintenance:

\subsubsection{Row Tracking}
\label{subsec:row_tracking}

Row tracking~\cite{row_track} assigns a unique identifier to each row at insertion time, and the identifier is preserved across row updates.
This enables efficient row-level lineage tracking and MV updates~\cite{katsis2015utilizing}.

\subsubsection{Change Data Feed (CDF)}

A change data feed (CDF) \cite{cdf} tracks row-level changes between table commits. Implemented as a table-valued function, a table's CDF returns one row per insertion or deletion, matching the source table's schema with an additional metadata column indicating the change type (as +1 or -1).

An \emph{effectivized changeset} (analogous to \emph{consolidation} in Differential Dataflow~\cite{mcsherry2013differential}) is the most compact representation of a table's CDF between two commits, canceling operations within the same changeset. For example, if a row was inserted and then deleted from a table, the net effect of these operations cancels. An effectivized changeset, however, can still contain multiple operations for a given row, e.g., an update is represented as a delete followed by an insert.

To effectivize a changeset, its rows are grouped by all columns to compute the sum of the change type column in each group. Then only rows with non-zero net changes are kept.

\subsubsection{Deletion Vectors}
Deletion vectors \cite{dvs} implement a merge-on-read strategy for Delta Lake~\cite{armbrust2020delta}: instead of rewriting entire Parquet~\cite{vohra2016apache} files on every change (copy-on-write), \texttt{DELETE}, \texttt{UPDATE}, and \texttt{MERGE} operations mark rows as removed without rewriting files. This reduces write amplification at the cost of additional merging during reads.

\subsubsection{Time Travel}
Each data manipulation language (DML) operation creates a new table version. Time travel \cite{time_travel} allows reading data from any historical version.

%% file: sections/ivm_definition.tex
\section{Incremental Refresh of Materialized Views}
\label{sec:problem_definition}

In this section, we start by formally describing the incremental view maintenance problem and then proceed to formally define Enzyme's recursive incremental plan constructions used across a variety of operators and query shapes.

\subsection{Incremental View Maintenance}
\label{subsec:ivm_def}

Consider an MV that stores the result of a query \( Q(T_1, \ldots, T_n) \) defined over
a set of base relations \( T_1, T_2, \ldots, T_n \).
When the contents of one or more base tables are modified, Enzyme aims to
automatically compute the corresponding change to the MV, denoted by \(\Delta Q\),
such that the refreshed view satisfies:
\[
Q(T_1', \ldots, T_n') = Q(T_1, \ldots, T_n) \oplus \Delta Q,
\]
where \(T_i'\) denotes the updated state of table \(T_i\), and
\(\oplus\) represents the application of the incremental update
(typically a combination of insertions and deletions)
to the existing MV state.
In general, for each base relation we define
\(\Delta T_i = T_i' - T_i\)
where "\(-\)" denotes the bag (multi-set)
difference,
and the incremental view maintenance problem is to compute \(\Delta Q\)
given \(Q\), the current base tables \(T_i\), and their corresponding changes \(\Delta T_i\).
The \(\Delta\) terms are concretely represented as relations with rows annotated with the extra change type column denoting insertions (+1) and deletions (-1).

The central challenge addressed by Enzyme is to perform this computation
\emph{automatically} for arbitrary SQL queries without user intervention.
That is, given any query \(Q\), Enzyme derives a correct and efficient
incremental maintenance plan to update the MV in response to
changes in its inputs, rather than recomputing the entire query from scratch.

A naive strategy for incrementalization is to exploit the identity
\(\Delta Q = Q(T_1', \ldots, T_n') - Q(T_1, \ldots, T_n)\).
That is, one can evaluate \(Q\) on both the previous and the updated states of the inputs,
and compute their difference to obtain \(\Delta Q\).
While this approach is conceptually straightforward and always correct for deterministic queries (\Cref{subsec:non_determinism} discusses why determinism matters here),
it requires evaluating \(Q\) twice, which is prohibitively expensive in most practical settings—often more costly
than simply recomputing the MV from scratch.
Nevertheless, this formulation provides the foundation for constructing
more efficient incrementalization strategies for complex queries,
as we discuss below.

\subsection{Operator-Level Delta Plan Construction}
\label{subsec:ivm_logic}

\begin{figure}[t]
    \centering
    \includegraphics[width=0.7\linewidth, trim={0cm 0.6cm 0cm 0.7cm}, clip]{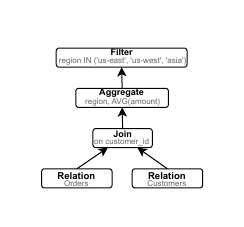}
    \caption{
    Simplified query plan for the MV query in \Cref{fig:running_example_mv}.
    }
    \label{fig:query_plan}
\end{figure}

To tackle the problem of efficiently incrementalizing queries, Enzyme operates
over the individual operators in the logical query plan of an MV rather than
treating the query as a monolithic whole.
Spark DataFrames and SQL queries can be expressed as trees of relational operators such as
\texttt{Project}, \texttt{Filter}, \texttt{Aggregate}, \texttt{Window}, \texttt{Join}, etc., where each operator transforms one or more
input relations into an output relation~\cite{spark_sql}.
\Cref{fig:query_plan} shows the simplified query plan for our
running example query.

Enzyme operates on these primitive operators and constructs
\emph{delta-plan fragments} corresponding to each operator, which are
then composed to maintain the result of the overall query~\cite{griffinL95, griffinK98_outerjoins}.
This decomposition works because for any operator node \(\phi\) in the query plan,
we can treat its children \(\psi_i\) as relations that may receive updates
\(\Delta \psi_{i}\).
Conceptually, each node behaves as if it materializes an intermediate relation.
Thus, incrementalization at the operator level reduces to defining a delta
transformation \(\Delta \phi\) such that
\[
\phi(\psi')
=
\phi(\psi \oplus \Delta \psi)
=
\phi(\psi)
\;\oplus\;
\Delta \phi.
\]
For simplicity, the above formulation assumes that \(\phi\) is a unary operator with
one child \(\psi\).
For operators with multiple children, the definition generalizes naturally by
taking deltas from all child nodes as inputs.

\begin{figure}
    \centering
    \includegraphics[width=\linewidth]{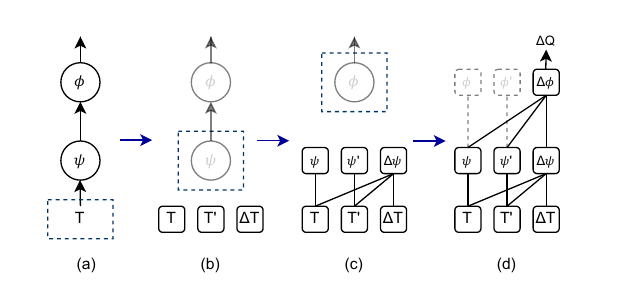}
    \caption{
    Illustration of operator-level delta plan composition.
    The incremental plan is constructed by traversing the query plan from the leaf nodes upward and progressively applying delta rules.
    At each operator node \(\psi\), three outputs are generated:
    \(\psi\) (the previous output), \(\psi'\) (the output over the updated input), and \(\Delta \psi\) (the change up to this node).
    These intermediate deltas are then used by parent operators, following the algebraic formulations in \Cref{subsec:ivm_logic},
    to compose the final incremental plan \(\Delta Q\).
    }
    \label{fig:delta_plan}
\end{figure}

In general, computing \(\Delta \phi\) may require access to any combination of \(\psi'\), \(\psi\),
and \(\Delta \psi\), as illustrated by the operator-level rules discussed later.
Although \(\psi'\) can be obtained by applying
\(\Delta \psi\) to the previous state \(\psi\), Enzyme propagates both
\(\psi\) and \(\psi'\) explicitly for efficiency, as reflected in the
algebraic transformations described later.
Once per-operator delta transformations are defined, they can be composed
bottom-up—following the structure of the query plan—to produce the final
incremental plan \(\Delta Q\), as illustrated in \Cref{fig:delta_plan}.

Below we outline the intuition behind incrementalizing several key operator types.
Many of these operator-level formulations have been extensively studied in prior work on
incremental view maintenance~\cite{griffinK98_outerjoins, griffinL95, gupta1993maintaining, quass96_aggregation}.

\paragraph{Projection (\( \pi \)) and Filter (\( \sigma_\theta \)).}
For projection and filter operators, the delta propagates directly through:
$\Delta(\pi_{a_1, \ldots, a_k}(T)) = \pi_{a_1, \ldots, a_k}(\Delta T)$ and
$\Delta(\sigma_\theta(T)) = \sigma_\theta(\Delta T)$.
Since projection and filter operate on each tuple independently,
their incremental update can be obtained by simply applying the same operator
to the input delta.

\paragraph{Aggregation (\( \mathcal{G}_{k, agg}\)).}
Aggregations are more complex because each output tuple summarizes
a group of input tuples. A naive recomputation
would re-aggregate all groups, but given \(\Delta T\), only those groups affected
by the changes need to be updated. This leads to the following formulation,
where \(\pi_{-}\) and \(\pi_{+}\) denote deletions and insertions
produced at this operator:
\[
\Delta(\mathcal{G}_{k, agg}(T)) =
\pi_{-}(\mathcal{G}_{k, agg}(T \ltimes_{k} \Delta T))
+
\pi_{+}(\mathcal{G}_{k, agg}(T' \ltimes_{k} \Delta T)).
\]
For example, for \( \mathcal{G}_{k, \texttt{SUM}(f)}(T) \),
it suffices to recompute the sums for only the groups (i.e., distinct \(k\) values)
that appear in \(\Delta T\).
This principle generalizes to a wide range of aggregation functions,
including \texttt{SUM}, \texttt{AVG}, \texttt{COUNT}, and \texttt{MEDIAN},
as long as the computation is deterministic (see \Cref{subsec:non_determinism} to
see how non-deterministic functions are handled).
As discussed later in~\Cref{subsec:ivm_special_cases}, Enzyme further extends this formulation
with specialized techniques for handling top-level and compositional aggregations.

\paragraph{Window.}
For window functions with a \texttt{PARTITION BY} clause,
the maintenance is analogous to that of aggregations above;
only the partitions affected by \(\Delta T\) require recomputation.

\paragraph{Joins.}
Join operators are among the most challenging to incrementalize.
The delta for an inner join \( L \bowtie R \) can be expressed as
$\Delta(L \bowtie R) = (\Delta L \bowtie R) + (L' \bowtie \Delta R)$~\cite{gupta1993maintaining}.

Outer joins can be maintained by conceptually decomposing them into an inner join
and one (for left- or right-outer) or two (for full-outer) anti-joins, yielding a formulation
similar to that proposed in prior work~\cite{griffinK98_outerjoins}.
Although this approach is theoretically correct and applies to all types of outer joins,
its performance degrades significantly when multiple joins are composed.
To address this, Enzyme employs specialized delta-construction techniques for
the most common join patterns, such as equi-joins and primary-foreign key joins.
We omit the full details of these constructions due to space constraints.

To illustrate how we combine the individual rules described above, we now present the delta plan
construction for our running example in \Cref{fig:running_example_mv}.
In the above notation, this query is expressed as
$\mathcal{G}_{\text{region},\, \texttt{AVG}(\text{amount})}
\big(
  \text{Customers} \bowtie_{\text{customer\_id}} \text{Orders}
\big)$.

We can now repeatedly apply the operator-level delta rules from above,
pushing the \(\Delta\) terms down to the leaf relations.
For brevity, we write \(C\) and \(O\) for \texttt{Customers} and \texttt{Orders},
respectively, and omit the predicates, and join and grouping keys.
\[
\begin{aligned}
\Delta Q
  &= \Delta ( \sigma (\mathcal{G}(C \bowtie O))) \\[-0.25em]
  &= \sigma (\Delta ( \mathcal{G}(C \bowtie O))) \\[-0.25em]
  &= \sigma (\pi_{-} ( \mathcal{G}(C \bowtie O) \ltimes \Delta (C \bowtie O))
     +
     \pi_{+} ( \mathcal{G}(C' \bowtie O') \ltimes \Delta (C \bowtie O))) \\[-0.25em]
  &= \sigma (\pi_{-} ( \mathcal{G}(C \bowtie O) \ltimes (C \bowtie \Delta O + \Delta C \bowtie O)) ~+ \\[-0.25em]
  &\qquad \pi_{+}( \mathcal{G}(C' \bowtie O') \ltimes (C \bowtie \Delta O + \Delta C \bowtie O)))
\end{aligned}
\]

\subsection{Applying Computed Changes}\label{sec:applying-changes}
Given a query \( Q \), we have seen how a corresponding \(\Delta Q\) plan can be derived
through the algebraic transformations discussed above.
These transformations are implemented in Enzyme as a semantic rewrite layer that generates executable delta plans.
To make the MV up to date, however, the computed changes must be
\emph{applied} to the MV's backing table.

To support efficient application of updates, Enzyme introduces an internal
\texttt{row\_id} column that uniquely identifies each output tuple at every stage
of the query plan.
At the leaf level, Enzyme relies on Delta Lake's row tracking
(see \Cref{subsec:row_tracking}) to provide stable, unique identifiers for each input row.
At higher operator levels, Enzyme deterministically constructs new identifiers that
are unique to the output of each operator.
For example, for a join operator, the \texttt{row\_id} is derived by combining
the row identifiers from the left and right inputs;
for an aggregation, the grouping keys serve as the identifier for each aggregated row.

Once each operator output is associated with a unique \texttt{row\_id},
the resulting \(\Delta Q\) can be used to determine which rows in the MV
should be deleted or inserted.
Enzyme first identifies the set of \texttt{row\_id} values corresponding to deletions
and removes them from the MV's backing table. It
then inserts the newly produced rows corresponding to insertions.
The precise Delta operations used to apply a change set to the MV's backing table
are discussed further in Sections~\ref{subsubsec:levearing_mv} and~\ref{sec:execution}.

\subsection{Handling Non-Determinism}
\label{subsec:non_determinism}
A largely overlooked aspect in the IVM literature is how the non-deterministic
behavior of certain operators or expressions can affect the correctness of incremental plan generation.
Most existing incrementalization techniques, including those discussed above, rely on the ability to deterministically reconstruct the previous output state of each operator given its prior inputs.

Consider the example query below, however:
\begin{lstlisting}[frame=none]
SELECT a, rand() AS r FROM T WHERE r > 0.5
\end{lstlisting}
When a row is deleted from \texttt{T}, a naive application of the rules above would evaluate
\texttt{rand()} for the deleted tuple to compute the corresponding delta.
But since \texttt{rand()} is non-deterministic, it will likely
produce a different value upon re-evaluation, and
as a result, the computed change would
no longer reflect the difference between the old and new states.
In this example, a tuple that initially satisfied the predicate
(\texttt{r > 0.5}) might later not during deletion,
incorrectly skipping its removal from the MV.

While functions such as \texttt{rand()} are explicitly non-deterministic,
non-determinism also arises in more subtle ways.
Examples include aggregate functions like
\texttt{collect\_set} or \texttt{collect\_list}, whose output order is
not guaranteed; floating-point aggregates like \texttt{SUM} or \texttt{AVG},
where arithmetic precision introduces non-deterministic variation;
and window functions with ordering clauses, where ties may be broken arbitrarily.

Enzyme mitigates these issues through semantic-preserving rewrites whenever
possible and when doing so does not introduce significant performance
overhead.
For example, Enzyme introduces an explicit local sort to enforce a
deterministic order for the output of each \texttt{collect\_set}.
Because these sorts are applied locally within each partition,
they do not require cross-node data shuffles and thus incur minimal overhead.

In cases where a deterministic rewrite is not possible---such as queries using
\currenttimestamp{}, whose evaluation depends on the current system time---Enzyme tries to employ specialized incrementalization
strategies tailored to the most common patterns observed in
production workloads. Some of these are discussed in~\Cref{subsec:ivm_special_cases}.
If neither a rewrite nor a specialized technique is possible, e.g., when
handling a non-deterministic UDF, Enzyme simply falls back to complete recomputation.

\subsection{Optimizations \& Special Cases}
\label{subsec:ivm_special_cases}

In \Cref{subsec:ivm_logic}, we described a general approach for constructing incremental plans for arbitrary queries.
This approach performs well across a wide range of query patterns but can be
suboptimal for certain query shapes.
Moreover, there are patterns that are not supported by the general formulation.
In this section, we outline several cases where Enzyme
leverages specialized logic to ensure correctness and efficiency.

These specializations are \emph{local} in nature:
each one replaces or refines the default delta rule for a specific operator
node in the query plan. The overall bottom-up composition procedure described
in \Cref{subsec:ivm_logic} remains unchanged. During plan construction,
Enzyme checks whether a node matches one of the specialized cases below
and, if so, substitutes the corresponding optimized rule in place of the
general formulation. The resulting delta fragments compose with those of
surrounding operators exactly as before.

\subsubsection{Temporal Filters} \label{subsec:temporal_filters}
As mentioned in~\Cref{subsec:non_determinism}, queries containing
\currenttimestamp{} (and related expressions such as
\currentdate{}) are non-deterministic and cannot be handled directly
by the general IVM formulation.
But a very common pattern observed in production workloads is the use of
these expressions to define a \emph{temporal filter}, so the MV contains only the most recent \(n\) days (or hours) of data.
As an example, suppose we update our running example MV to compute each region's 30-day rolling average sales by adding a filter on the \texttt{Orders} relation: \lstinline|WHERE date >= current_date() - INTERVAL 30 DAYS|.

To handle such cases, Enzyme includes specialized logic to capture the values of time-dependent functions used in filter predicates in the previous and current refreshes. It then uses the previous and current timestamps to calculate the source rows entering and leaving the temporal window.

Formally, consider a filter \( f(t) \) that depends on a time variable \( t \)
and is applied to a base table \( T \): \( \sigma_{f(t)}(T) \).
Let \( f(t=\text{prev}) \) and \( f(t=\text{curr}) \) denote the filter predicate with $t$ set to the previous and current refresh timestamps, respectively.
The delta for this temporal filter can be computed as:
\[
\begin{aligned}
\Delta(\sigma_{f(t)}(T)) &= \pi_{-}\!\left(\sigma_{\neg f(t=\text{curr}) \,\wedge\, f(t=\text{prev})}(T)\right) \;+\; \\
&\quad\pi_{+}\!\left(\sigma_{\neg f(t=\text{prev}) \,\wedge\, f(t=\text{curr})}(T)\right) \;+\; \\
&\quad\sigma_{f(t=\text{curr})}(\Delta T)
\end{aligned}
\]
Intuitively, the first term removes source rows that have fallen out of temporal window;
the second term adds rows that have now moved into the window;
and the third term applies any changes that remain valid under the
current temporal predicate.

This incrementalization formula is actually a special case of handling modifications to an MV's definition. For instance, handling a change in a filter from \lstinline|WHERE col = 'x'|
to \lstinline|WHERE col = 'y'| can be handled similarly if we capture the parameters used in the previous and current refreshes (values `x' and `y' in this case), although this approach may not be efficient in general.

\subsubsection{Leveraging Materialized Data}
\label{subsubsec:levearing_mv}
The general incrementalization formulation described earlier relies on the
pre-state (\(T\)), post-state (\(T'\)), and the computed changes (\(\Delta T\))
of base tables to derive the delta of a query.
But for queries whose top-level operator is an aggregation or a window
function, we can further optimize the process by exploiting the fact that the
previous output of the query is already materialized in the MV itself.
This allows Enzyme to avoid recomputing the previous state of the output from scratch.

Consider again our running example MV from \Cref{fig:running_example_mv}.
In the general formulation, Enzyme would compute the changeset resulting
from the join of the \texttt{Customers} and \texttt{Orders} tables (using  \(\Delta \text{Customers}\) and \(\Delta \text{Orders}\)).
It would then use this to identify the distinct
\texttt{region} values (i.e., grouping keys) requiring updates and recompute both
the previous and current \texttt{avg\_order\_amount} values for those grouping keys
to determine \(\Delta Q\).
However, as the MV already stores the previous aggregate results,
Enzyme can instead simply delete all rows whose grouping keys appear in
computed join changeset and then append the newly computed aggregate values.
This approach eliminates the need to re-evaluate the previous aggregate state
entirely and applies analogously to queries with top-level window functions.

For certain aggregation functions, such as \texttt{SUM} and \texttt{COUNT},
Enzyme can push this optimization further by maintaining incremental
adjustments to the aggregate value, rather than recomputing it from scratch.
To support this, Enzyme maintains additional metadata---such as the count of
tuples per group---if it is not already present among the aggregated columns.

Aggregations such as \texttt{AVG} and \texttt{STDDEV} are internally
\emph{decomposed} into simpler aggregates (\texttt{SUM} and \texttt{COUNT})
that can be incrementally updated using these merge-based delta adjustments~\cite{gupta1993maintaining}. As a result, for the running example query, Enzyme runs the aggregation only on
the computed join changeset to compute the delta of the aggregate and then uses a
\texttt{MERGE INTO} operation to update the MV's rows in place.

\subsubsection{Partition Overwrite}
The general formulation computes changes at the granularity of individual
rows; it determines the precise set of tuples to delete and insert.
In some cases, however, this fine-grained approach can be suboptimal.
When both the input tables and the MV are partitioned on the same
columns, it is sometimes more efficient to compute changes at a coarser granularity
by refreshing entire partitions rather than individual rows.
This \emph{partition-overwrite} strategy is applicable when the
query does not contain operations that span multiple partition keys.

For example, consider tables and an MV all partitioned by a
\texttt{date} column.
If the query does not include aggregations or window functions that operate
across multiple \texttt{date} values, the system can safely refresh data at the
partition level.
In such cases, Enzyme overwrites the affected partitions instead of
computing and applying row-level deltas, which can significantly reduce
execution overhead associated with computing these deltas and applying
changes at the finer granularity.
Enzyme's cost model (discussed in \Cref{subsec:cost_model}) determines when to use partition overwrite versus
row-granularity refresh based on the relative cost of two incremental strategies.

%% file: sections/architecture.tex
\section{System Architecture}

Building a production-ready incremental view maintenance system presents challenges beyond the theoretical foundations. Although the algorithmic techniques for incrementalizing individual operators are well-established, productionizing these techniques requires addressing practical concerns including query plan stability, compatibility with existing systems, and robustness across diverse workloads. This section describes Enzyme's system architecture and the key components that enable reliable incremental refresh in a production environment.

\begin{figure}[b]            
  \centering
  \includegraphics[width=.98\linewidth]{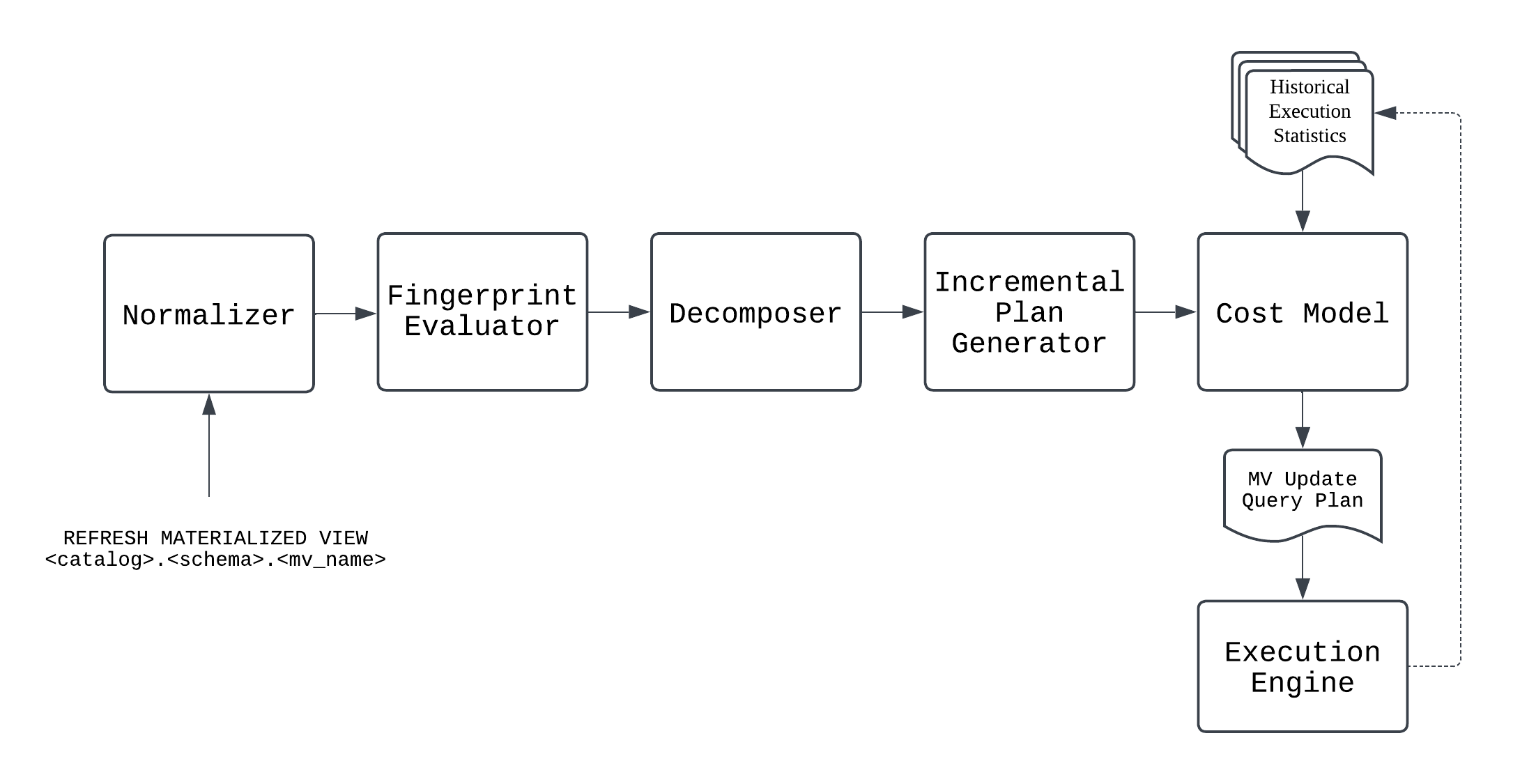} 
  \caption{Enzyme transforms a query into an incremental execution in six phases: normalization, fingerprinting, decomposition, plan generation, costing, and execution.}
  \label{fig:components}
\end{figure}

Enzyme's architecture comprises several interconnected components that work together to provide end-to-end incremental view maintenance. The system processes user queries through six main stages: normalization, fingerprinting, decomposition, incremental plan generation, costing, and execution. Figure~\ref{fig:components} shows a high-level diagram of Enzyme's architecture. 

\subsection{Normalization}
The main entry point for Enzyme is a Catalyst logical plan~\cite{spark_sql} representing the MV definition. This allows Enzyme to support all Spark front-ends, including SQL and the entire DataFrame API (e.g., for PySpark). 

The first step of Enzyme's processing is \emph{normalization}, which 
serves two critical purposes: 
(1) simplifying query plans to enable simpler incremental plan construction, and
(2) serving as the first step to ensure that logically equivalent queries with minor cosmetic differences 
produce identical fingerprints (discussed in \Cref{subsec:fingerprinting}). 
As shown below, the normalizer produces a simplified plan that lies between 
the analyzed and fully optimized plans.
\begin{figure}[H]
  \centering
  \includegraphics[width=.98\linewidth]{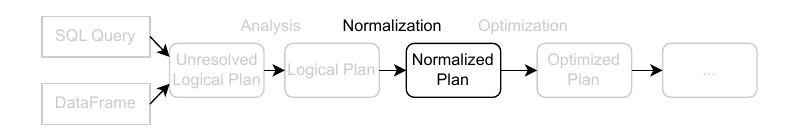}
  \label{fig:normalized_plan}
\end{figure}

Query plans arriving at Enzyme are logical plans and thus often contain redundancies and complexities that complicate the 
writing of incrementalization transformations. 
The normalizer applies a comprehensive set of simplification rules, including inlining common table expressions (CTEs), merging and simplifying filter predicates, collapsing unnecessary projections, and eliminating redundant operations.

Enzyme requires a normalized plan (instead of the fully optimized plan) because
some of Spark’s Catalyst optimizations~\cite{spark_sql}, while beneficial for execution, obscure or
eliminate information essential for incremental processing. 
For example, timestamp substitution and empty-relation propagation may rewrite or
remove expressions that are semantically meaningful for computing incremental updates. 
Normalizing the logical plan allows Enzyme to preserve the semantic
structure necessary for correct incremental refresh, while still producing a
canonical representation for fingerprinting and subsequent transformations.

\subsection{Query Fingerprinter}
\label{subsec:fingerprinting}

To ensure correctness, Enzyme must detect when users modify an MV's definition and trigger complete recomputation on the next refresh when changes occur. While catalog-based change tracking could address some scenarios, several factors make this approach insufficient for our purposes.

First, Enzyme must handle MVs defined in both SQL and Python, where source code and logical definitions can diverge. A common pattern observed in production is the use of meta-programming in Python to define complex pipelines with repetitive structures. Such cases make static, catalog-based tracking particularly difficult: source code can change without altering an MV's semantics (e.g., a code refactor) or an MV's definition can change without any source code modifications (e.g., changes in external parameters). Second, fingerprints derived from the normalized query plan naturally capture transitive changes through dependency chains---for example, when an MV references a view that itself references another view. Finally, by operating on the normalized plan, the fingerprinter ignores semantics-preserving changes, avoiding unnecessary recomputation.

\paragraph{Fingerprint Construction}
Beyond the normalization performed in the previous stage, the fingerprinter applies additional canonicalization steps to eliminate semantics-independent differences. These include standardizing the order of commutative operators, such as joins and unions, and the order of operands within commutative expressions, as well as ensuring consistent assignment of expression identifiers across sub-expressions.

For UDFs implemented in Python, fingerprinting is more involved because their semantics cannot be easily analyzed through static inspection alone. Enzyme addresses this by incorporating UDF signatures into the MV's fingerprint. The system tracks not only the UDF definition (using the Python bytecode) but also its transitive dependencies, ensuring that changes to imported libraries or nested function calls trigger appropriate recomputation. This comprehensive approach enables safe incremental processing even in environments with complex and evolving user code.

\paragraph{Fingerprint Stability}
Fingerprinting the normalized plan, however, also introduces a key challenge: maintaining fingerprint stability across system upgrades and algorithmic improvements. When Enzyme introduces new incremental techniques or optimizes existing ones, the fingerprints of affected MVs may change even though the user's query does not. Naively, this would trigger costly recomputation for all affected MVs during system upgrades.

Enzyme addresses this through multi-versioning of each MV's fingerprint. Before deploying changes affecting fingerprints, we introduce a new fingerprint version that maps the updated plan representation to the existing fingerprint, preserving continuity across the upgrade. Old fingerprint versions are retired alongside the corresponding MV data versions once they are no longer needed. This approach minimizes recomputation while ensuring all MVs can leverage improved incremental techniques.

\subsection{Decomposition \& Technique Enablers}
\label{sec:decomposition}

The split architecture of MVs at Databricks, comprising a backing Delta table and a top-level view, provides crucial flexibility for incremental computation. This design allows Enzyme to modify the internal representation of an MV while maintaining a consistent external interface for users.

Enzyme leverages this flexibility through \emph{technique enablers}, which are transformations applied to MV definitions to make incremental computation possible or more efficient. The most fundamental enabler is the decomposition of aggregate functions into incrementally computable components~\cite{gupta1993maintaining}. For example, an MV containing $\AVG(x)$ is internally augmented with additional columns storing $\SUM(x)$ and $\COUNT(*)$, enabling efficient incremental updates when new data arrives. The decomposer also performs targeted expression rewriting, such as converting semantically equivalent expressions like FIRST to MIN, where the ordering guarantees make them equivalent.
Another critical enabler is the explicit propagation of aggregate and window keys to the top-level output of the query plan. This transformation exposes grouping and partition keys in the plan output, which can be exploited by our incremental plan generator (discussed below).
Finally, the technique enablers handle the construction of the unique row identifiers described in \Cref{sec:applying-changes}.

\subsection{Incremental Plan Generator}
The incremental plan generator transforms a normalized and enabled query plan into one that computes the MV's incremental changes based on the rules defined in \Cref{subsec:ivm_logic}. 
This component employs a recursive visitor pattern that traverses the query plan bottom-up, producing at each node a representation of how to compute changes to that node's output given changes to its inputs.

\paragraph{Composable Delta Plan Construction.} At the core of the generator is the concept of a delta plan, which comprises three pieces: the pre-state (the data before changes), the post-state (the data after changes), and the delta (the actual change representation with insertion/deletion markers). As the visitor traverses each operator in the query plan, it transforms the delta plans from child nodes into a delta plan for the current node by applying the operator-level delta rules described in \Cref{subsec:ivm_logic}. 
Finally, the completed traversal yields an incremental plan to refresh the MV.

\paragraph{Effectivization.} Enzyme's plan generator selectively performs effectivization for changesets consumed by intermediate operators in the plan. This is technically optional but can be useful to reduce downstream computation when the change volume is large and contains redundant pairs.

\paragraph{Further Optimizations.} As discussed in \Cref{subsubsec:levearing_mv}, for MVs with top-level aggregations or window functions, Enzyme applies an additional optimization. Rather than computing the full changeset with complete before-and-after images of affected rows, the generator recognizes that only the aggregation or partition keys are needed to identify which rows to delete, along with the after-image values for insertion. This eliminates the need to materialize the full pre-state of affected groups, significantly reducing the cost when the aggregation or window is expensive to compute.

\input{sections/cost_model}

\subsection{Refresh Execution}
\label{sec:execution}

While Enzyme's planning components operate on logical query plans, the actual execution of these plans occurs within Apache Spark, with Enzyme as a client. This architectural decision allows Enzyme to benefit from ongoing improvements to the Spark optimizer and execution engine without re-implementing physical planning or execution logic. 
Enzyme operates on Spark logical plans, specifying transformations and actions that are then optimized and executed by Spark's Catalyst optimizer~\cite{spark_sql} and execution engine~\cite{behm2022photon,xue2024adaptive}.

\paragraph{Spark Configuration and Optimization.}
Before executing an incremental plan, Enzyme tunes several Spark configuration 
parameters to optimize performance based on expected workload characteristics. 
Because Enzyme constructs the incremental plans itself and has access to 
historical change statistics, it has additional information that can be used to 
steer the optimizer toward more informed decisions.
For example, when a delta term participates in a join, Enzyme may use historical 
changeset sizes to hint the optimizer toward appropriate join strategies 
(e.g., broadcast vs.\ shuffle). 
Similarly, Enzyme may adjust broadcast thresholds dynamically based on the 
changeset sizes observed in previous refreshes.

For complex plans involving multiple aggregations or joins, Enzyme may selectively cache intermediate results that will be reused multiple times within the same refresh operation. This is particularly valuable when computing both the pre- and post-states of inputs for aggregate changesets, where caching the input changeset avoids redundant computation.

\paragraph{Choosing Update Mechanisms.} Given an incremental plan produced by the plan generator, Enzyme applies the computed changes to the MV's backing table. It selects between two mechanisms based on the changeset characteristics and query structure.

\texttt{REPLACE WHERE}~\cite{replacew} performs an atomic delete-then-insert operation. Since it semantically applies all deletions before any insertions, the final changeset must be effectivized to avoid incorrect results. Otherwise, a row appearing in both the deletion and insertion sets could introduce duplicate or stale rows. The execution layer is responsible for adding a final effectivization operation to un-effectivized \texttt{REPLACE WHERE} incremental plans.

For certain MVs with top-level aggregates, Enzyme uses an optimized \texttt{MERGE INTO}~\cite{mergeinto} formulation to update materialized results in place by combining existing values with incremental deltas (see~\Cref{subsubsec:levearing_mv}).

\paragraph{Transactional Provenance Updates.} Enzyme maintains provenance metadata for each MV that records the query fingerprint, source table versions, and other information necessary to validate consistency and detect definition changes. This provenance is stored as part of the MV's metadata and must remain consistent with its data. Enzyme ensures this consistency by committing provenance changes in the same Delta transaction~\cite{armbrust2020delta} that applies the data changes. Delta's ACID transaction semantics guarantee that either both the data and provenance are updated atomically, or neither update is visible. This prevents the MV from entering an inconsistent state where the data reflects one set of source versions while the provenance metadata references others.

%% file: sections/cost_model.tex
\subsection{Cost Model}
\label{subsec:cost_model}

\begin{figure}[t]            
  \centering
  \includegraphics[width=.98\linewidth]{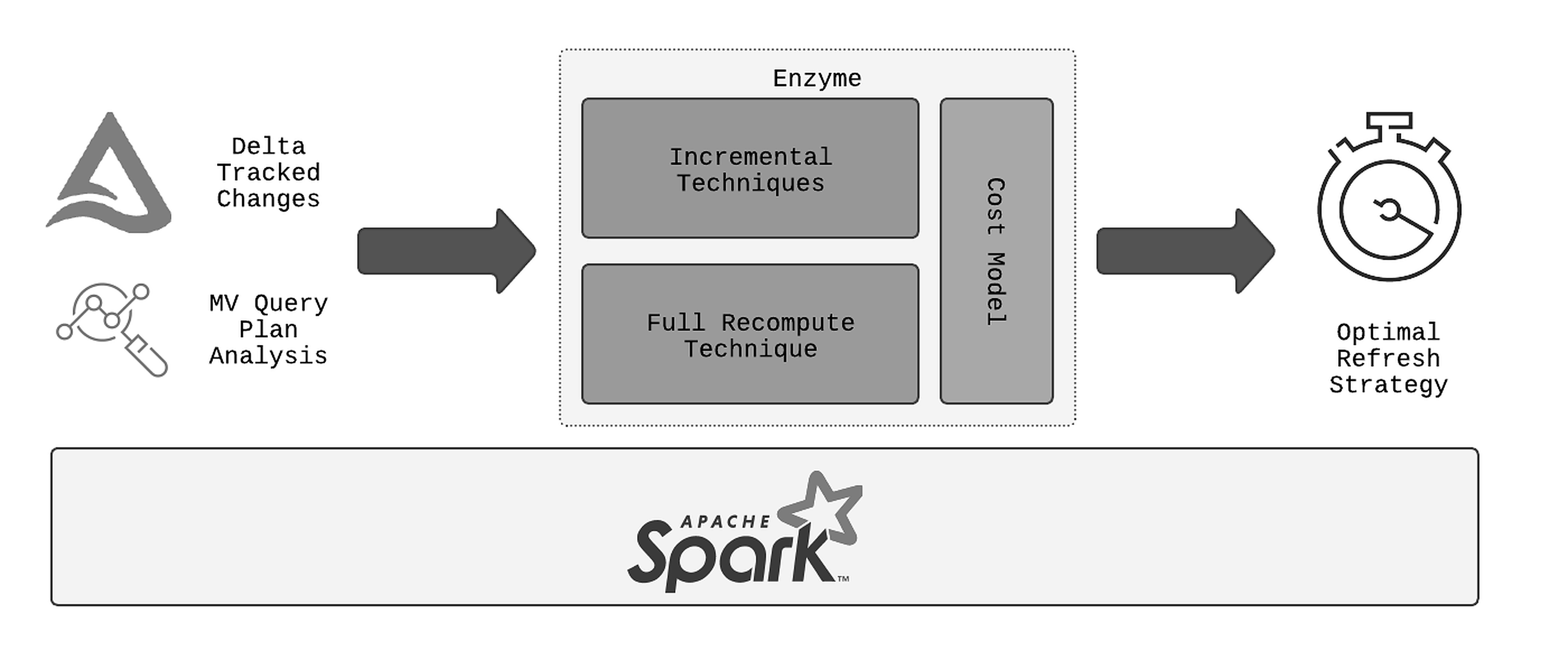} 
  \caption{Enzyme selects between incremental and full recomputation techniques based on estimated execution cost grounded by historical executions.}
  \label{fig:planner}
\end{figure}

An effective IVM engine requires sophisticated cost modeling. Multiple logically equivalent query plans exist for updating an MV, yet cost models in most online analytical processing (OLAP) databases lack sufficient accuracy to reliably choose between them~\cite{leis2015good}. Enzyme's cost model addresses this gap by estimating the end-to-end execution cost of data transformations across layers of the medallion architecture. For each MV, it selects the optimal refresh strategy—incremental update or full recomputation (\Cref{fig:planner})—by modeling the total refresh cost as the sum of executor CPU times across key physical operators, including joins, aggregates, window functions, shuffles, file scans, and file writes.

Enzyme leverages execution profiles from prior refreshes to ground cost estimates in empirical data. For each operator in a refresh plan under consideration, the system identifies structurally similar historical executions through normalized physical plan matching and derives expected CPU time from observed execution metrics. This approach implicitly captures runtime phenomena such as data skew, spilling behavior, and dynamic pruning effects recorded in prior executions. When no historical information is available, the model falls back to default parameters derived from large-scale query processing logs.

This costing framework, with its historical execution feedback loop, allows Enzyme to function as an adaptive, data-driven system capable of accurately modeling diverse query structures and workload characteristics while maintaining operational reliability at scale. The cost model serves as a fundamental component that opens opportunities for automatically determining which intermediate MVs would benefit customer workloads by trading additional storage for compute savings~\cite{agrawal2000automated, ahmed2020automated}. Due to its complexity and significance, a comprehensive treatment of Enzyme's cost model is reserved for future work.

%% file: sections/lessons.tex
\section{Operational Experience \& Lessons}

Deploying an IVM engine in a production data platform requires rigorous attention to correctness, performance, and reliability. This section describes key operational practices and lessons learned from running Enzyme at scale.

\paragraph{Ensuring Correctness.}
Enzyme employs multiple complementary strategies to validate correctness. The system leverages a Random Query Generator (RQG) framework to generate random schemas, synthetic data, and SQL queries covering complex combinations of different operators. For each generated query, Enzyme creates an MV, applies randomized changes to source tables, and validates that the incremental refresh produces identical results to complete recomputation. To complement synthetic testing, Enzyme leverages an A/B correctness testing system for production workloads that compares MV contents to their definition queries. This dual approach catches subtle correctness issues that manifest only under specific data distributions or timing conditions not covered by synthetic tests.

\paragraph{Reliability Through Fallback.}
Beyond correctness validation, ensuring reliability requires graceful degradation in the face of bugs and other issues. Enzyme implements automatic fallback to complete recomputation when planning encounters exceptions or incorrectly generates an invalid incremental plan. The system monitors these exceptions closely, as they often reveal bugs in the underlying query engine. Enzyme also integrates with the Health Mediated Release framework in Databricks, which automatically rolls back to the previously deployed version when a refresh fails after a system upgrade. The complexity of incremental plans exercises execution paths rarely encountered by typical user queries, making Enzyme an effective stress test for uncovering optimizer and execution bugs in Spark itself.

\paragraph{Performance Optimization Challenges.}
Operating Enzyme at scale revealed several performance challenges requiring careful attention. While Enzyme leverages Spark's query optimizer when submitting incremental plans for execution, several scenarios required considering alternative plan formulations. For example, when dynamic file pruning optimizations failed to activate, we adjusted Enzyme to use an equivalent incremental formulation that leverages semijoins to provide more explicit pruning. Additionally, some incremental plans caused out-of-memory failures in the driver during optimization due to plan complexity. Enzyme addresses this by adding rules to the normalizer to simplify plans, or by reformulating incremental formulas to be more tractable for the optimizer.

\paragraph{Pipeline-Aware Cost Modeling}
A key insight from deploying Enzyme's cost model is that optimization decisions must consider the entire graph of dependent MVs rather than optimizing each view in isolation. Even when an upstream MV would benefit from complete recomputation, performing incremental refresh may be preferable because it generates a smaller change feed that dramatically reduces refresh costs for downstream dependent MVs. This holistic approach can reduce total system cost despite being locally suboptimal for individual MVs.

\paragraph{Fingerprint Stability Challenges.}
Maintaining fingerprint stability over time required addressing spurious fingerprint changes by leveraging multi-versioning. Changes to the normalizer, for instance, should not change fingerprints and invalidate existing MVs. Similarly, the upgrade from Scala 2.12 to 2.13 potentially introduced many spurious changes due to differences in how Scala objects were hashed. The multi-versioned fingerprinter has been critical for allowing Enzyme to handle such cases without triggering unnecessary recomputation.

%% file: sections/experimental_study.tex
\section{Experimental Study}

\begin{figure}[b]            
  \centering
  \includegraphics[width=.9\linewidth]{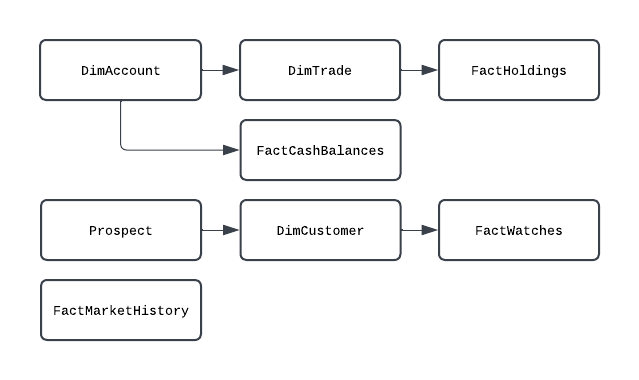} 
  \caption{Dependency Graph of MVs in TPC-DI.}
  \Description{Dependency Graph of MVs in the TPC-DI implementation.}
  \label{fig:tpc_di_components}
\end{figure}

We evaluate the effectiveness of Enzyme in performing incremental updates compared to both full recomputation and a state of the art IVM system. Our experiments aim to demonstrate three key outcomes: (1) Enzyme delivers superior performance and efficiency over full recomputation across scale factors, (2) it outperforms existing commercial alternatives in incremental update workloads, and (3) its cost model reliably identifies when incrementalization is most beneficial.

\subsection{Experimental Setup}

In the experiments for this paper, we use the TPC-DI benchmark \cite{poess2014tpc} as our workload. TPC-DI is well-suited for evaluating ETL and incremental processing systems because it includes built-in incremental updates, realistic operational domains, and a mix of append-only and CDC-style changes.

\subsubsection{Data and Query Workload} The TPC-DI benchmark models end-to-end data warehousing across multiple operational domains—Customer, Account, Trade, Company, Security, Financial, and Market—integrated into a unified analytical schema. It consists of three batches: a large historical load covering over two years of data, followed by two single-day incremental batches. Although TPC-DI does not prescribe an ingestion method, we adopt a streaming ingestion layer, consistent with modern systems and aligned with the Databricks Medallion Architecture. This design preserves the benchmark’s temporal semantics while cleanly separating record acquisition (append or CDC processing, deduplication, surrogate key generation) from analytical derivation.

Operational datasets such as TradeHistory, DailyMarket, and Financial are modeled as append-only streaming tables, while entity datasets, such as Customer, Account, Company, and Security, are represented as CDC streaming tables that maintain current state through merges.

On top of this ingestion layer, analytical datasets—including DimCustomer, DimAccount, DimSecurity, FactHoldings, and FactMarketHistory—are expressed as MVs that perform multi-table joins, aggregations, and enrichment, corresponding to the Silver/Gold layers of the Medallion Architecture.

To ensure parity, we use the same logical definitions for both Enzyme and the compared cloud system. Our study focuses on evaluating the IVM performance of each system, measuring compute efficiency during updates. Datasets that do not receive changes in a given batch are excluded from evaluation.

\begin{figure*}[ht]
    \centering
    \includegraphics[width=0.95\textwidth]{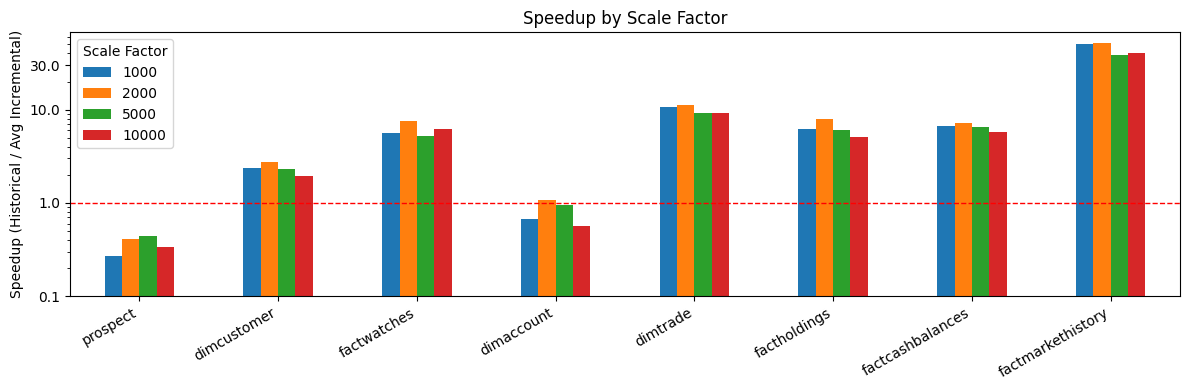}
    \caption{Enzyme performance overview on TPC-DI benchmark. Note that the y-axis is log scale.}
    \Description{Bar chart denoting Enzyme's performance on the TPC-DI benchmark. Logarithmic scale is used on the y-axis.}
    \label{fig:databricks_updates}
\end{figure*}

\subsubsection{Enzyme Setup}
We create a Spark Declarative Pipeline (SDP) with our ingestion layer defined as Streaming Tables and the analytical datasets defined as MVs. A Databricks Serverless cluster is used as the underlying compute. Data is generated for all 3 batches with the DIGen tool provided by TPC-DI and stored into a staging location. Before each batch, we copy the data for the batch to an ingestion location for the pipeline and execute an update of the pipeline.

Data is generated at the 1000, 2000, 5000, and 10,000 scale factors and executed with the same batch sequencing. 
In all scale factors, we used a serverless autoscaling pipeline backed by AWS c7g.2xlarge executor instances (8 CPU cores, 16 GiB memory each). To reduce variance and improve reproducibility, we capped the number of executors at 64, which does not fully reflect production capacity.

\subsubsection{\vendor{} Setup}
We compare Enzyme against a commercial Cloud Vendor. Due to licensing restrictions, we refer to it simply as \vendor{}. In the \vendor{} setup, we use raw tables for ingestion and defined IVM objects on top of them. For each batch, we copy the data into the raw tables, and trigger a refresh of the IVM objects in topological order to ensure functional equivalence with Enzyme’s automated pipeline updates.

TPC-DI data at the 10,000 scale factor was used to test the \vendor{} performance. The underlying compute used 64 c7g.2xlarge executor nodes.

\subsection{Experimental Results}
To account for differences in hardware and resource configurations, we report relative speedups—expressed as a multiple of the runtime of a full recomputation of the same query. Although the cost models may select full recomputation for certain queries, we report incremental performance results for all queries to present a comprehensive comparison.

These results show that:
\begin{itemize}
    \item Enzyme successfully supports and incrementalizes \textbf{100\% of the TPC-DI workloads}, demonstrating broad applicability across diverse dataset types.
    \item The Enzyme cost model made the \textbf{optimal decision in 7 out of 8 cases} with one false negative selection.
    \item When incrementalization is applied, Enzyme achieves \textbf{better performance than full recomputation in 6 out of 8 datasets}, confirming the effectiveness of its incremental execution model.
    \item The observed performance benefits are \textbf{consistent across larger scale factors}, indicating that Enzyme’s efficiency scales with data volume.
\end{itemize}

\subsubsection{Comparison with Full Recomputes in Databricks}\label{full_recompute_comparison}

Figure \ref{fig:databricks_updates} shows the relative speedup for each scale factor tested in Databricks for each dataset. We summarize our findings as an in-depth breakdown of the overhead and query characteristics is outside the scope of this paper.

Across the benchmark, six out of eight datasets exhibit substantial speedups when executed incrementally compared to full recomputation. The speedups remain roughly constant across scale factors, demonstrating Enzyme’s ability to maintain efficiency and scalability as data volume increases.

\textbf{FactMarketHistory} is a computationally intensive workload that computes 52-week highs and lows for each stock symbol based on roughly two years of historical data. Each refresh requires scanning one year of records per symbol, leading to high memory pressure and increased disk spilling at larger scale factors. This explains the dip in speedup between scale factors 2000 and 5000.

\textbf{Prospect} performs poorly under incremental refresh because a large fraction of the dataset is updated in each batch. The column SK\_RecordDateID represents the most recent date a prospective customer appears in the input. Since more than 95\% of prospects persist across batches, nearly all rows in the dataset are rewritten during each refresh to update this column. In such cases, identifying and applying fine-grained row-level updates incurs more overhead than simply performing a full table refresh.

\textbf{DimAccount} is a lightweight workload that completes in under a minute even at the largest scale. Small datasets like this are more sensitive to transient overheads—including cache contention and scheduling delays—caused by concurrently running queries. Although DimAccount itself shows limited speedup, incrementalizing it remains important because it enables more efficient incremental updates for its downstream consumers.

\begin{figure*}[ht]
    \centering
    \includegraphics[width=0.95\textwidth]{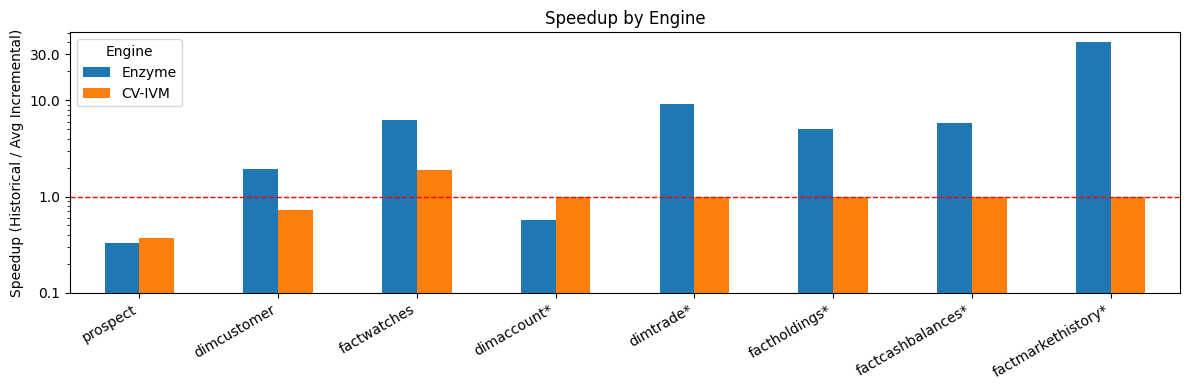}
    \caption{Enzyme performance compared to a leading cloud vendor. Note that the y-axis is log scale.}
    \vspace{0.3em}
    {\footnotesize *Speedups were reported as 1 for queries that the vendor did not support incrementalizing.}
    \Description{Bar chart comparing Enzyme performance against CV-IVM on each MV. If CV-IVM did not incrementalize the MV, a default speedup of 1.0 is used.}
    \label{fig:engine_updates}
\end{figure*}

\subsubsection{Comparison with \vendor{}}
Figure \ref{fig:engine_updates} shows the relative speedup of Enzyme compared to \vendor{}, tested on the 10,000 scale factor. Although the static cost model in the \vendor{} offering did not choose to incrementalize any of the datasets, we overrode the cost model to perform incremental updates where possible.

When setting up the workload in \vendor{}, DimAccount and FactMarketHistory were not able to be refreshed incrementally due to them containing unsupported operators. Although DimTrade, FactHoldings, FactCashBalances did not have unsupported operators, they were forced to perform full refreshes because their upstream dependencies were not incrementalized.

Of the three datasets that did incrementalize, we observed better performance with FactWatches, and the incremental refresh in \vendor{} regressed in Prospect and DimCustomer. Prospect likely suffered for the same reason discussed in section~\ref{full_recompute_comparison}. However it was not apparent why DimCustomer regressed with \vendor{}.

\subsubsection{Cost model evaluation on TPC-DI} Across the TPC-DI workload, the Enzyme cost model made the right decision for 7 out of the 8 datasets, representing an accuracy of 87.5\%.

\textbf{Prospect}, as discussed in Section~\ref{full_recompute_comparison}, is more efficiently handled by a full recompute—consistent with the cost model’s choice.

\textbf{FactCashBalances} was excluded from incrementalization due to operator nesting detected by the model; however, empirical results show that its incremental refresh actually outperforms a full recompute.

\textbf{DimAccount} was chosen to be incremental. Although it did not show significant gains when incrementalized, it was globally optimal as it helped the performance of downstream datasets.

The cost model used by \vendor{} makes static decisions based on the user query, and did not choose to incrementalize any of the datasets in the benchmark.

\begin{figure}[ht]
    \centering
    \begin{subfigure}{\linewidth}
        \centering
        \includegraphics[width=.8\linewidth]{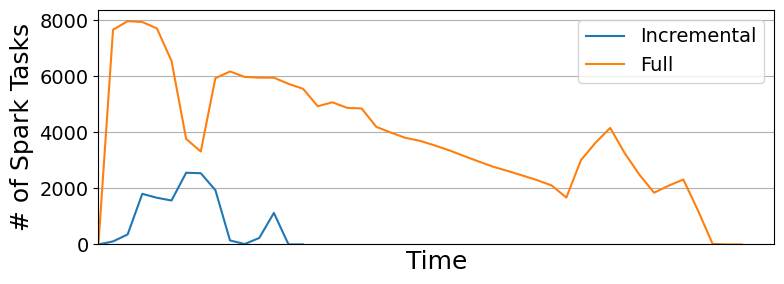}
        \caption{Task history during historical vs incremental execution}
        \label{fig:task_history}
    \end{subfigure}

    \vspace{0.5em}

    \begin{subfigure}{\linewidth}
        \centering
        \includegraphics[width=.8\linewidth]{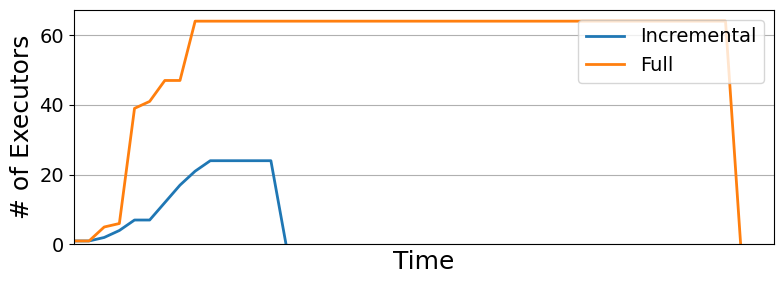}
        \caption{Executor history during historical vs incremental execution}
        \label{fig:executor_history}
    \end{subfigure}

    \caption{Autoscaling takes advantage of smoother task load during incremental runs to scale down and improve efficiency.}
    \Description{Autoscaling takes advantage of smoother task load during incremental runs to scale down and improve efficiency. The graphs show the number of Spark tasks and executors in the cluster for full and incremental refreshes.}
    \label{fig:history_metrics}
\end{figure}

\subsubsection{Autoscaling outcomes} Spark Declarative Pipelines in Databricks support automatic scaling, dynamically adjusting compute resources based on workload demand.
As shown in Figure \ref{fig:task_history}, the incremental runs exhibit a curve in the number of submitted Spark tasks with smaller spikes, indicating reduced data volume and more consistent scheduling.
Consequently, as illustrated in Figure \ref{fig:executor_history}, the cluster requires fewer executors overall, since the smoother and lower task load eliminates the sharp peaks in parallelism seen during full recomputes. This leads to more efficient resource utilization and lower overall compute cost, as the cluster scales only to the level necessary for the incremental workload.

%% file: sections/related_works.tex
\section{Related Work}

Incremental view maintenance (IVM) has been a long-standing problem in databases. Early work \cite{blakeley1986efficiently, griffinL95, griffinK98_outerjoins, gupta1993maintaining} developed first-order delta rules that propagate inserts, deletes, and updates through relational algebra, with follow-ups covering both set and bag semantics; nested and correlated subqueries remained difficult and only partially addressed by later theory.

Higher-order IVM advanced the state of the art. DBToaster \cite{dbtoaster11cidr, koch2014dbtoaster} applies recursive finite differencing~\cite{dbtoaster10pods}, materializing not only the view but also higher-order deltas, trading additional state for substantially lower update costs on join-heavy queries. 
Differential Dataflow \cite{mcsherry2013differential} generalizes incremental computation using partially ordered timestamps and consolidation, while DBSP \cite{budiu2022dbsp} recasts databases as streams over commutative groups with integration/differentiation operators; the formalisms are closely related.

Several prior works have proposed algorithms and data structures for 
incrementalizing specific classes of queries. 
For example, DYN~\cite{dyn_sigmod17} and IDYN~\cite{idyn_sigmod19, idyn_vldb18} target acyclic conjunctive queries (CQs) with 
equality and inequality join predicates, respectively; 
AJU~\cite{aju_sigmod_20} focuses on incremental maintenance of acyclic foreign-key joins; 
and RPAI~\cite{rpai_sigmod22} addresses correlated nested aggregate queries.

A complementary line exploits keys. ID-based IVM \cite{katsis2015utilizing} represents change sets as identifiers rather than full tuples, often avoiding base-table lookups on the update path and yielding large speedups on multi-join workloads, subject to key/foreign-key assumptions. Its modular compilation—ID inference, rule instantiation/composition, and algebraic minimization—reduces both work and state. 

Production systems span a wide spectrum. Traditional MVs in Redshift \cite{redshift}, BigQuery \cite{bigquery}, and Postgres-style \cite{postgresmv} warehouses target query acceleration with periodic refresh.
Modern streaming MVs (e.g., Materialize \cite{mcsherry2022materialize}, RisingWave \cite{wang2022sneak}) are typically deployed as long-running dataflows that continuously process input streams, as opposed to scheduled batch refreshes, enabling low-latency maintenance.
Cloud warehouses bridge paradigms: Snowflake Streams and Dynamic Tables \cite{akidau2023s} pair CDC with scheduled/target-freshness maintenance. Databricks provides MVs refreshed manually or on a schedule; streaming tables which provide append-oriented ingestion with exactly-once processing in managed pipelines. Both surfaces integrate with Delta Lake Change Data Feed for row-level CDC. Streaming tables inherit structured streaming semantics such as watermarks for state bounding and exactly-once processing when using appropriate sources/sinks.

Materialization strategy has been studied from view selection (AND–OR DAGs; multi-query optimization \cite{mistry2001materialized}) to recursive maintenance (semi-naive evaluation \cite{green2013datalog}; DRed and counting variants \cite{gupta1993maintaining}). These classics underpin today’s systems: DBToaster’s aggressive intermediate caches exemplify the state-vs-latency trade-off, while differential/DBSP frameworks formalize streaming execution and consolidation.

Enzyme delivers a fully automated, built-in approach for Databricks MVs capable of handling arbitrary compositions of complex SQL operators including aggregations, joins, and window functions without user intervention. Its cost-based optimization layer intelligently selects refresh strategies across entire pipelines of MVs, automatically choosing between incremental updates and full recomputation based on historical execution profiles and empirical feedback. On the other hand, streaming tables are designed to seamlessly process and store incoming data streams in real time. Unlike traditional batch tables that require explicit refreshes or data overwrites, streaming tables automatically ingest new data as it arrives, maintaining a continuously updated, up-to-date view. This design makes streaming tables inherently incremental. The combination of  streaming tables and MVs creates a unified, powerful framework that elegantly addresses both batch and streaming use cases. 

%% file: sections/conclusion.tex
\section{Conclusion}

Enzyme is a state-of-the-art incremental view maintenance engine that synthesizes decades of foundational academic research into an industrial-grade system, addressing the critical challenge of maintaining MV consistency in modern data lakehouse architectures as source data evolves. Built atop Apache Spark primitives, it achieves exceptional operator coverage and seamlessly handles complex operator compositions including aggregations, joins, and window functions. Among the few production-grade IVM engines, Enzyme demonstrates superior performance, validating both its scalability and practical effectiveness across thousands of real-world pipelines. Successfully bridging the gap between theoretical advances and operational demands, Enzyme is an essential optimization system for data processing in the Lakehouse era.

%% file: references.bib
@online{redshift,
  author       = {{Amazon Web Services}},
  title        = {Materialized Views in {Amazon Redshift}},
  year         = {2024},
  url          = {https://docs.aws.amazon.com/redshift/latest/dg/materialized-view-overview.html},
  lastaccessed = {November 1, 2025}
}

@online{autocdc,
  author       = {{Databricks}},
  title        = {The {AUTO CDC} {APIs}: Simplify Change Data Capture with Pipelines},
  year         = {2025},
  url          = {https://docs.databricks.com/aws/en/ldp/cdc},
  lastaccessed = {November 1, 2025}
}

@online{bigquery,
  author       = {{Google}},
  title        = {Introduction to Materialized Views},
  year         = {2025},
  url          = {https://cloud.google.com/bigquery/docs/materialized-views-intro},
  lastaccessed = {November 1, 2025}
}

@online{postgresmv,
  author       = {{PostgreSQL Global Development Group}},
  title        = {Materialized Views},
  year         = {2025},
  url          = {https://www.postgresql.org/docs/current/rules-materializedviews.html},
  lastaccessed = {November 1, 2025}
}

@online{mergeinto,
  author       = {{Databricks}},
  title        = {{MERGE INTO} ({Delta Lake SQL} Reference)},
  year         = {2025},
  url          = {https://docs.databricks.com/aws/en/sql/language-manual/delta-merge-into},
  lastaccessed = {November 1, 2025}
}

@inproceedings{rpai_sigmod22,
  author    = {Abeysinghe, Supun and He, Qiyang and Rompf, Tiark},
  title     = {Efficient Incrementialization of Correlated Nested Aggregate Queries
               using Relative Partial Aggregate Indexes ({RPAI})},
  booktitle = {Proceedings of the ACM International Conference on Management of Data
               (SIGMOD '22)},
  pages     = {136--149},
  publisher = {ACM},
  year      = {2022},
  doi       = {10.1145/3514221.3517889}
}

@inproceedings{agrawal2000automated,
  author    = {Agrawal, Sanjay and Chaudhuri, Surajit and Narasayya, Vivek R.},
  title     = {Automated Selection of Materialized Views and Indexes in {SQL} Databases},
  booktitle = {Proceedings of the 26th International Conference on Very Large Data Bases
               (VLDB '00)},
  pages     = {496--505},
  publisher = {Morgan Kaufmann},
  address   = {Cairo, Egypt},
  year      = {2000}
}

@article{ahmed2020automated,
  author    = {Ahmed, Rafi and Bello, Randall and Witkowski, Andrew and Kumar, Praveen},
  title     = {Automated Generation of Materialized Views in {Oracle}},
  journal   = {Proceedings of the VLDB Endowment},
  volume    = {13},
  number    = {12},
  pages     = {3046--3058},
  year      = {2020},
  publisher = {VLDB Endowment}
}

@article{akidau2023s,
  author    = {Akidau, Tyler and Barbier, Paul and Cseri, Istvan and Hueske, Fabian
               and Jones, Tyler and Lionheart, Sasha and Mills, Daniel
               and Pauliukevich, Dzmitry and Probst, Lukas and Semmler, Niklas
               and Sotolongo, Dan and Zhang, Boyuan},
  title     = {What's the Difference? {Incremental} Processing with Change Queries
               in {Snowflake}},
  journal   = {Proceedings of the ACM on Management of Data},
  volume    = {1},
  number    = {2},
  pages     = {1--27},
  year      = {2023},
  publisher = {ACM},
  doi       = {10.1145/3589776}
}

@article{armbrust2020delta,
  author    = {Armbrust, Michael and Das, Tathagata and Sun, Liwen and Yavuz, Burak
               and Zhu, Shixiong and Murthy, Mukul and Torres, Joseph
               and van Hovell, Herman and Ionescu, Adrian and {\L}uszczak, Alicja
               and {\'{S}}wi{\k{a}}tkowski, Micha{\l} and Szafra{\'{n}}ski, Micha{\l}
               and Li, Xiao and Ueshin, Takuya and Mokhtar, Mostafa and Boncz, Peter
               and Ghodsi, Ali and Paranjpye, Sameer and Senster, Pieter
               and Xin, Reynold and Zaharia, Matei},
  title     = {{Delta Lake}: High-Performance {ACID} Table Storage over Cloud Object Stores},
  journal   = {Proceedings of the VLDB Endowment},
  volume    = {13},
  number    = {12},
  pages     = {3411--3424},
  year      = {2020},
  doi       = {10.14778/3415478.3415560}
}

@inproceedings{spark_sql,
  author    = {Armbrust, Michael and Xin, Reynold S. and Lian, Cheng and Huai, Yin
               and Liu, Davies and Bradley, Joseph K. and Meng, Xiangrui
               and Kaftan, Tomer and Franklin, Michael J. and Ghodsi, Ali
               and Zaharia, Matei},
  title     = {{Spark SQL}: Relational Data Processing in {Spark}},
  booktitle = {Proceedings of the ACM International Conference on Management of Data
               (SIGMOD '15)},
  pages     = {1383--1394},
  publisher = {ACM},
  year      = {2015},
  doi       = {10.1145/2723372.2742797}
}

@inproceedings{armenatzoglou2022amazon,
  author    = {Armenatzoglou, Nikos and Basu, Sanuj and Bhanoori, Naga and Cai, Mengchu
               and Chainani, Naresh and Chinta, Kiran and Govindaraju, Venkatraman
               and Green, Todd J. and Gupta, Monish and Hillig, Sebastian
               and Hotinger, Eric and Leshinksy, Yan and Liang, Jintian
               and McCreedy, Michael and Nagel, Fabian and Pandis, Ippokratis
               and Parchas, Panos and Pathak, Rahul and Polychroniou, Orestis
               and Rahman, Foyzur and Saxena, Gaurav and Soundararajan, Gokul
               and Subramanian, Sriram and Terry, Doug},
  title     = {{Amazon Redshift} Re-invented},
  booktitle = {Proceedings of the ACM International Conference on Management of Data
               (SIGMOD '22)},
  pages     = {2205--2217},
  publisher = {ACM},
  year      = {2022}
}

@inproceedings{behm2022photon,
  author    = {Behm, Alexander and Palkar, Shoumik and Agarwal, Utkarsh
               and Armstrong, Timothy and Cashman, David and Dave, Ankur
               and Greenstein, Todd and Hovsepian, Shant and Johnson, Ryan
               and Sai Krishnan, Arvind and Leventis, Paul and Luszczak, Ala
               and Menon, Prashanth and Mokhtar, Mostafa and Pang, Gene
               and Paranjpye, Sameer and Rahn, Greg and Samwel, Bart
               and van Bussel, Tom and van Hovell, Herman and Xue, Maryann
               and Xin, Reynold and Zaharia, Matei},
  title     = {Photon: A Fast Query Engine for Lakehouse Systems},
  booktitle = {Proceedings of the ACM International Conference on Management of Data
               (SIGMOD '22)},
  pages     = {2326--2339},
  publisher = {ACM},
  year      = {2022},
  doi       = {10.1145/3514221.3526054}
}

@inproceedings{bello1998materialized,
  author    = {Bello, Randall G. and Dias, Karl and Downing, Alan
               and Feenan, James J. and Finnerty, James L. and Norcott, William D.
               and Sun, Harry and Witkowski, Andrew and Ziauddin, Mohamed},
  title     = {Materialized Views in {Oracle}},
  booktitle = {Proceedings of the 24th International Conference on Very Large Data Bases
               (VLDB '98)},
  pages     = {659--664},
  publisher = {Morgan Kaufmann},
  address   = {New York, NY, USA},
  year      = {1998}
}

@inproceedings{blakeley1986efficiently,
  author    = {Blakeley, Jos{\'e} A. and Larson, Per-{\AA}ke and Tompa, Frank Wm.},
  title     = {Efficiently Updating Materialized Views},
  booktitle = {Proceedings of the ACM International Conference on Management of Data
               (SIGMOD '86)},
  pages     = {61--71},
  publisher = {ACM},
  address   = {Washington, DC, USA},
  year      = {1986},
  doi       = {10.1145/16894.16861}
}

@article{budiu2022dbsp,
  author    = {Budiu, Mihai and Chajed, Tej and McSherry, Frank
               and Ryzhyk, Leonid and Tannen, Val},
  title     = {{DBSP}: Automatic Incremental View Maintenance for Rich Query Languages},
  journal   = {Proceedings of the VLDB Endowment},
  volume    = {16},
  number    = {7},
  pages     = {1601--1614},
  year      = {2023},
  publisher = {VLDB Endowment},
  doi       = {10.14778/3587136.3587137}
}

@inproceedings{chandra2025unity,
  author    = {Chandra, Ramesh and Chen, Haogang and Matharu, Ray and Cai, Sarah
               and Chen, Jeff and Dutta, Priyam and Ghita, Bogdan and Greenstein, Todd
               and Holla, Gopal and Huang, Peng and Huo, Yuchen and Ionescu, Adrian
               and Ispas, Adriana and Januschowski, Tim and Karajgaonkar, Vihang
               and Leone, Stefania and Lewis, David and Li, Andrew and Li, Nong
               and Lian, Cheng and Link, Stephen and Lu, Qing and Ma, Yesheng
               and Pettitt, Chris and Prabhakaran, Vijayan and Raducanu, Bogdan
               and Rong, Kyle and Roome, Paul and Shetty, Samarth and Smith, Sean
               and Sun, Xiaotong and Tang, Yuyuan and Wen, Weitao and Xia, Lei
               and Zeng, Junlin and Zhang, Ben and Xin, Reynold and Zaharia, Matei},
  title     = {{Unity Catalog}: Open and Universal Governance for the Lakehouse and Beyond},
  booktitle = {Companion of the ACM International Conference on Management of Data
               (SIGMOD '25)},
  pages     = {310--322},
  publisher = {ACM},
  year      = {2025}
}

@online{time_travel,
  author       = {{Databricks}},
  title        = {Introducing {Delta} Time Travel for Large Scale Data Lakes},
  year         = {2019},
  url          = {https://www.databricks.com/blog/2019/02/04/introducing-delta-time-travel-for-large-scale-data-lakes.html},
  lastaccessed = {November 1, 2025}
}

@online{row_track,
  author       = {{Databricks}},
  title        = {Use Row Tracking for {Delta} Tables},
  year         = {2024},
  url          = {https://docs.databricks.com/aws/en/delta/row-tracking},
  lastaccessed = {November 1, 2025}
}

@online{replacew,
  author       = {{Databricks}},
  title        = {Selectively Overwrite Data with {Delta Lake}},
  year         = {2025},
  url          = {https://docs.databricks.com/aws/en/delta/selective-overwrite},
  lastaccessed = {November 1, 2025}
}

@online{cdf,
  author       = {{Databricks}},
  title        = {Use {Delta Lake} Change Data Feed on {Databricks}},
  year         = {2025},
  url          = {https://docs.databricks.com/aws/en/delta/delta-change-data-feed},
  lastaccessed = {November 1, 2025}
}

@online{dvs,
  author       = {{Databricks}},
  title        = {What Are Deletion Vectors?},
  year         = {2025},
  url          = {https://docs.databricks.com/aws/en/delta/deletion-vectors},
  lastaccessed = {November 1, 2025}
}

@article{goldstein2001optimizing,
  author    = {Goldstein, Jonathan and Larson, Per-{\AA}ke},
  title     = {Optimizing Queries Using Materialized Views: A Practical, Scalable Solution},
  journal   = {ACM SIGMOD Record},
  volume    = {30},
  number    = {2},
  pages     = {331--342},
  year      = {2001},
  publisher = {ACM}
}

@article{green2013datalog,
  author    = {Green, Todd J. and Huang, Shan Shan and Loo, Boon Thau
               and Zhou, Wenchao},
  title     = {Datalog and Recursive Query Processing},
  journal   = {Foundations and Trends in Databases},
  volume    = {5},
  number    = {2},
  pages     = {105--195},
  year      = {2013},
  publisher = {Now Publishers},
  doi       = {10.1561/1900000017}
}

@article{griffinK98_outerjoins,
  author    = {Griffin, Timothy and Kumar, Bharat},
  title     = {Algebraic Change Propagation for Semijoin and Outerjoin Queries},
  journal   = {ACM SIGMOD Record},
  volume    = {27},
  number    = {3},
  pages     = {22--27},
  year      = {1998}
}

@inproceedings{griffinL95,
  author    = {Griffin, Timothy and Libkin, Leonid},
  title     = {Incremental Maintenance of Views with Duplicates},
  booktitle = {Proceedings of the ACM International Conference on Management of Data
               (SIGMOD '95)},
  pages     = {328--339},
  publisher = {ACM},
  address   = {San Jose, CA, USA},
  year      = {1995},
  doi       = {10.1145/223784.223849}
}

@inproceedings{gupta1993maintaining,
  author    = {Gupta, Ashish and Mumick, Inderpal Singh and Subrahmanian, V. S.},
  title     = {Maintaining Views Incrementally},
  booktitle = {Proceedings of the ACM International Conference on Management of Data
               (SIGMOD '93)},
  pages     = {157--166},
  publisher = {ACM},
  address   = {Washington, DC, USA},
  year      = {1993},
  doi       = {10.1145/170035.170066}
}

@inproceedings{dyn_sigmod17,
  author    = {Idris, Muhammad and Ugarte, Mart{\'{\i}}n and Vansummeren, Stijn},
  title     = {The Dynamic {Yannakakis} Algorithm: Compact and Efficient Query
               Processing Under Updates},
  booktitle = {Proceedings of the ACM International Conference on Management of Data
               (SIGMOD '17)},
  pages     = {1259--1274},
  publisher = {ACM},
  year      = {2017},
  doi       = {10.1145/3035918.3064027}
}

@article{idyn_vldb18,
  author    = {Idris, Muhammad and Ugarte, Mart{\'{\i}}n and Vansummeren, Stijn
               and Voigt, Hannes and Lehner, Wolfgang},
  title     = {Conjunctive Queries with Inequalities Under Updates},
  journal   = {Proceedings of the VLDB Endowment},
  volume    = {11},
  number    = {7},
  pages     = {733--745},
  year      = {2018}
}

@article{idyn_sigmod19,
  author    = {Idris, Muhammad and Ugarte, Mart{\'{\i}}n and Vansummeren, Stijn
               and Voigt, Hannes and Lehner, Wolfgang},
  title     = {Efficient Query Processing for Dynamically Changing Datasets},
  journal   = {ACM SIGMOD Record},
  volume    = {48},
  number    = {1},
  pages     = {33--40},
  year      = {2019}
}

@inproceedings{katsis2015utilizing,
  author    = {Katsis, Yannis and Ong, Kian Win and Papakonstantinou, Yannis
               and Zhao, Kevin Keliang},
  title     = {Utilizing {IDs} to Accelerate Incremental View Maintenance},
  booktitle = {Proceedings of the ACM International Conference on Management of Data
               (SIGMOD '15)},
  pages     = {1985--2000},
  publisher = {ACM},
  year      = {2015}
}

@inproceedings{dbtoaster11cidr,
  author    = {Kennedy, Oliver and Ahmad, Yanif and Koch, Christoph},
  title     = {{DBToaster}: Agile Views for a Dynamic Data Management System},
  booktitle = {Proceedings of the 5th Biennial Conference on Innovative Data Systems
               Research (CIDR '11)},
  pages     = {284--295},
  publisher = {www.cidrdb.org},
  year      = {2011}
}

@inproceedings{dbtoaster10pods,
  author    = {Koch, Christoph},
  title     = {Incremental Query Evaluation in a Ring of Databases},
  booktitle = {Proceedings of the Twenty-Ninth ACM SIGMOD-SIGACT-SIGART Symposium
               on Principles of Database Systems (PODS '10)},
  pages     = {87--98},
  publisher = {ACM},
  year      = {2010},
  doi       = {10.1145/1807085.1807100}
}

@article{koch2014dbtoaster,
  author    = {Koch, Christoph and Ahmad, Yanif and Kennedy, Oliver and Nikolic, Milos
               and N{\"o}tzli, Andres and Lupei, Daniel and Shaikhha, Amir},
  title     = {{DBToaster}: Higher-Order Delta Processing for Dynamic, Frequently
               Fresh Views},
  journal   = {The VLDB Journal},
  volume    = {23},
  number    = {2},
  pages     = {253--278},
  year      = {2014},
  publisher = {Springer},
  doi       = {10.1007/s00778-013-0348-4}
}

@article{leis2015good,
  author    = {Leis, Viktor and Gubichev, Andrey and Mirchev, Atanas and Boncz, Peter
               and Kemper, Alfons and Neumann, Thomas},
  title     = {How Good Are Query Optimizers, Really?},
  journal   = {Proceedings of the VLDB Endowment},
  volume    = {9},
  number    = {3},
  pages     = {204--215},
  year      = {2015},
  publisher = {VLDB Endowment},
  doi       = {10.14778/2850583.2850594}
}

@inproceedings{mcsherry2022materialize,
  author    = {McSherry, Frank},
  title     = {Materialize: A Platform for Building Scalable Event Based Systems},
  booktitle = {Proceedings of the 16th ACM International Conference on Distributed
               and Event-Based Systems (DEBS '22)},
  pages     = {3},
  publisher = {ACM},
  year      = {2022}
}

@inproceedings{mcsherry2013differential,
  author    = {McSherry, Frank and Murray, Derek Gordon and Isaacs, Rebecca
               and Isard, Michael},
  title     = {Differential Dataflow},
  booktitle = {Proceedings of the 6th Biennial Conference on Innovative Data Systems
               Research (CIDR '13)},
  publisher = {www.cidrdb.org},
  year      = {2013}
}

@inproceedings{mistry2001materialized,
  author    = {Mistry, Hoshi and Roy, Prasan and Sudarshan, S. and Ramamritham, Krithi},
  title     = {Materialized View Selection and Maintenance Using Multi-Query Optimization},
  booktitle = {Proceedings of the ACM International Conference on Management of Data
               (SIGMOD '01)},
  pages     = {307--318},
  publisher = {ACM},
  year      = {2001}
}

@article{poess2014tpc,
  author    = {Poess, Meikel and Rabl, Tilmann and Jacobsen, Hans-Arno
               and Caufield, Brian},
  title     = {{TPC-DI}: The First Industry Benchmark for Data Integration},
  journal   = {Proceedings of the VLDB Endowment},
  volume    = {7},
  number    = {13},
  pages     = {1367--1378},
  year      = {2014},
  publisher = {VLDB Endowment}
}

@inproceedings{quass96_aggregation,
  author    = {Quass, Dallan},
  title     = {Maintenance Expressions for Views with Aggregation},
  booktitle = {Proceedings of the Workshop on Materialized Views: Techniques and
               Applications (VIEWS '96)},
  pages     = {110--118},
  year      = {1996}
}

@incollection{vohra2016apache,
  author    = {Vohra, Deepak},
  title     = {Apache {Parquet}},
  booktitle = {Practical {Hadoop} Ecosystem: A Definitive Guide to
               {Hadoop}-Related Frameworks and Tools},
  pages     = {325--335},
  publisher = {Springer},
  year      = {2016}
}

@inproceedings{aju_sigmod_20,
  author    = {Wang, Qichen and Yi, Ke},
  title     = {Maintaining Acyclic Foreign-Key Joins under Updates},
  booktitle = {Proceedings of the ACM International Conference on Management of Data
               (SIGMOD '20)},
  pages     = {1225--1239},
  publisher = {ACM},
  year      = {2020}
}

@inproceedings{wang2022sneak,
  author    = {Wang, Yanghao and Liu, Zhi},
  title     = {A Sneak Peek at {RisingWave}: A Cloud-Native Streaming Database},
  booktitle = {Proceedings of the 16th ACM International Conference on Distributed
               and Event-Based Systems (DEBS '22)},
  pages     = {190--193},
  publisher = {ACM},
  year      = {2022},
  doi       = {10.1145/3524860.3543284}
}

@article{xue2024adaptive,
  author    = {Xue, Maryann and Bu, Yingyi and Somani, Abhishek and Fan, Wenchen
               and Liu, Ziqi and Chen, Steven and van Hovell, Herman and Samwel, Bart
               and Mokhtar, Mostafa and Korlapati, Rk and Lam, Andy and Ma, Yunxiao
               and Ercegovac, Vuk and Li, Jiexing and Behm, Alexander and Li, Yuanjian
               and Li, Xiao and Krishnamurthy, Sriram and Shukla, Amit
               and Petropoulos, Michalis and Paranjpye, Sameer
               and Xin, Reynold and Zaharia, Matei},
  title     = {Adaptive and Robust Query Execution for Lakehouses at Scale},
  journal   = {Proceedings of the VLDB Endowment},
  volume    = {17},
  number    = {12},
  pages     = {3947--3959},
  year      = {2024},
  publisher = {VLDB Endowment}
}

@inproceedings{zilio2004recommending,
  author    = {Zilio, Daniel C. and Zuzarte, Calisto and Lightstone, Sam and Ma, Wenbin
               and Lohman, Guy M. and Cochrane, Roberta and Pirahesh, Hamid
               and Colby, Latha S. and Gryz, Jarek and Alton, Eric
               and Liang, Dongming and Valentin, Gary},
  title     = {Recommending Materialized Views and Indexes with the {IBM DB2}
               Design Advisor},
  booktitle = {Proceedings of the International Conference on Autonomic Computing
               (ICAC '04)},
  pages     = {180--187},
  publisher = {IEEE},
  year      = {2004}
}
